\documentclass[twoside]{dis04}
\def\runauthor{Leonid Gladilin}
\def\shorttitle{Heavy Flavours: experimental summary}
\begin{document}

\title{Heavy Flavours: experimental summary}

\author{Leonid Gladilin}

\address{DESY, ZEUS experiment, Notkestr. 85, 22607 Hamburg, Germany\\
E-mail: gladilin@mail.desy.de }

\maketitle

\abstracts{
The experimental talks presented in the working group D ``Heavy
Flavours''
of the DIS04 workshop
are summarised.
New and recently updated results from Tevatron, HERA, LEP, B-factories
and neutrino experiments are discussed. 
}

\section{Introduction}

Production and hadronisation of heavy quarks
is one of the most attractive laboratories for QCD prediction tests.
There is an active interplay between experimental analyses
and theoretical developments in this area.
The aim of the working group was to discuss
the relevant
experimental and theoretical results which become available or
updated after the previous DIS workshop.
The theoretical talks are summarised in~\cite{kretzer}
New and recently updated results from Tevatron, HERA, LEP, B-factories
and neutrino experiments are discussed in this note. 

\section{New resonances}

An observation of a narrow resonance 
decaying to $D^{*\pm}p^\mp$
has been reported by the H1 collaboration~\cite{h1ch5q}.
Fig.~1 shows the $D^{*\pm}p^\mp$
invariant-mass distributions
in DIS with $Q>1\,$GeV$^2$ and in photoproduction.
A fit of the signal in DIS yielded $50.6\pm11.2$ signal events,
the mass of $3099\pm3({\rm stat.})\pm5({\rm syst.})\,$MeV
and the Gaussian width of $12\pm3({\rm stat.})\,$MeV,
compatible with the experimental resolution.
A signal with compatible mass and width
was also observed in photoproduction.
The observed resonance was reported to contribute roughly $1\%$
to the total $D^{*\pm}$ production rate
in the kinematic region studied.
This resonance can be considered as a candidate for
the charmed pentaquark state, $\Theta^0_c = uudd{\bar c}$.

The observation of the H1 collaboration has been challenged
by the ZEUS collaboration~\cite{zeusch5q}.
Using a larger sample of $D^{*\pm}$ mesons,
ZEUS observed no signature of the narrow resonance in
the $M(D^{*\pm}p^\mp)$ spectra shown in Fig.~2.
The fake Gaussian signals shown in Fig.~\ref{fig:zeusch5q}
have the mass and the width of the H1 resonance and
contain numbers of events roughly estimated assuming
the resonance contributes $1\%$ to the visible $D^{*\pm}$ production rate.
The ZEUS data constrain the uncorrected fraction of $D^{*\pm}$ mesons
originating from $\Theta^0_c$ decays to be well below $1\%$.

The incompatibility between the H1 and ZEUS results on
the charmed pentaquark
should be clarified in further studies. The negative result on
the $\Theta^0_c$ search in $Z^0$ decays has been reported by the ALEPH
Collaboration in this conference~\cite{alephch5q}.
Reports from other collaborations are expected soon.

\begin{figure}[!ht]
\vspace*{-1.5cm}
\begin{center}
\centerline{\epsfxsize=2.9in\epsfbox{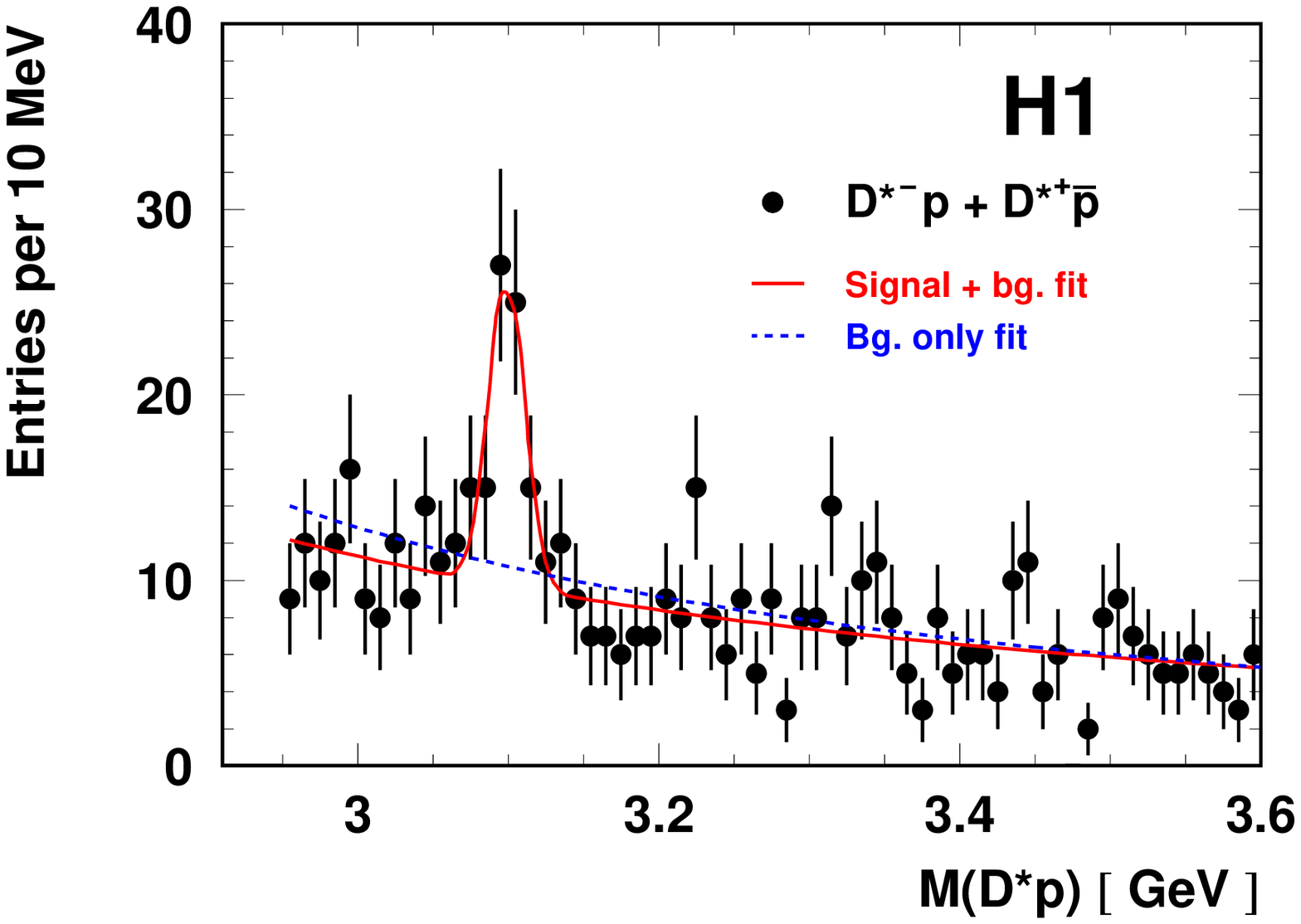}
\hspace*{-1.0cm}
\epsfxsize=2.9in\epsfbox{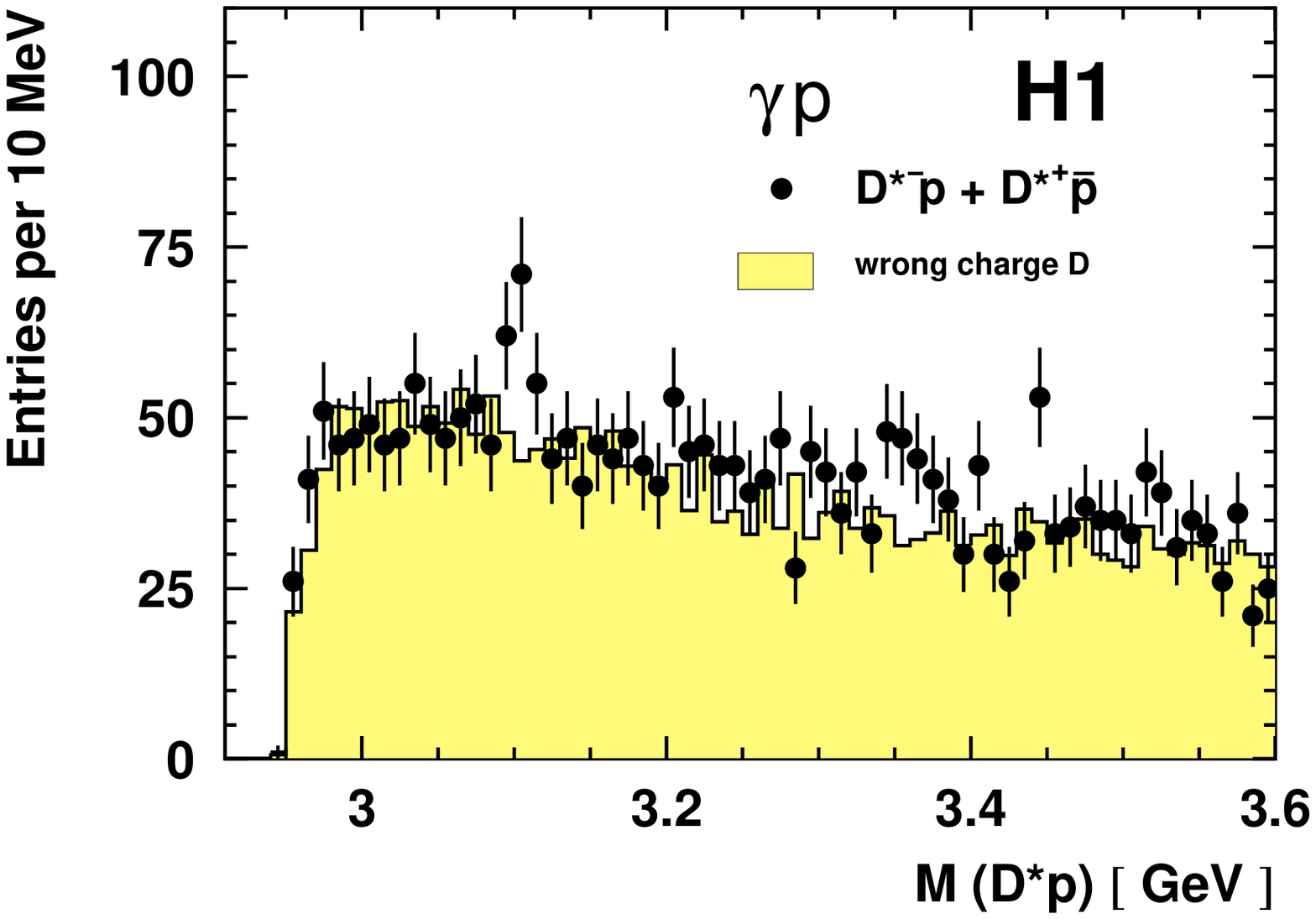}}
\caption[*]{
The distribution of $M(D^{*\pm}p^\mp)$
obtained by the H1 collaboration
in DIS (left) and photoproduction (right).
}
\end{center}
\label{fig:h1ch5q}
\end{figure}

\begin{figure}[!h]
\vspace*{-1.0cm}
\begin{center}
\centerline{\epsfxsize=2.8in\epsfbox{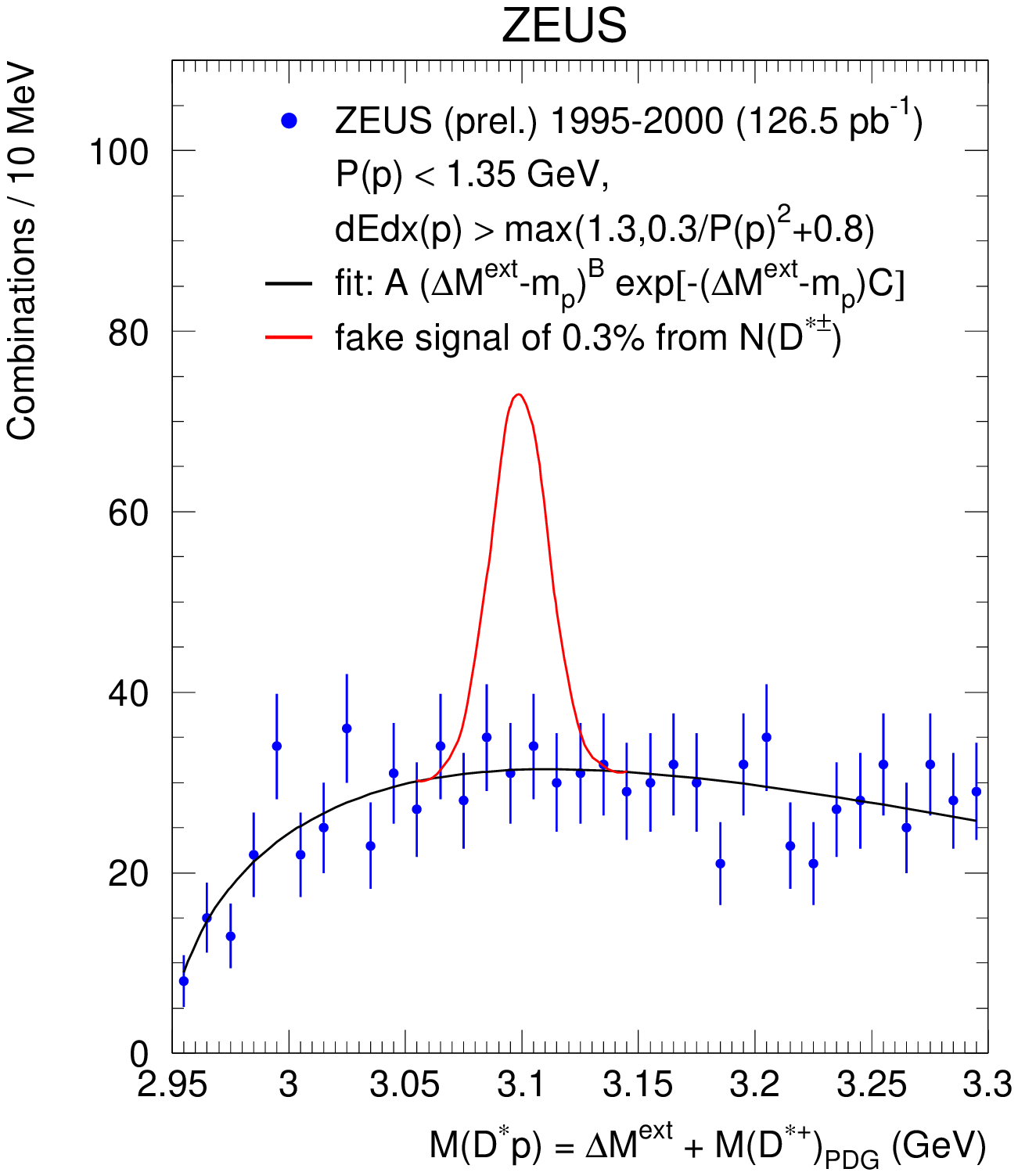}
\hspace*{-1.0cm}
\epsfxsize=2.8in\epsfbox{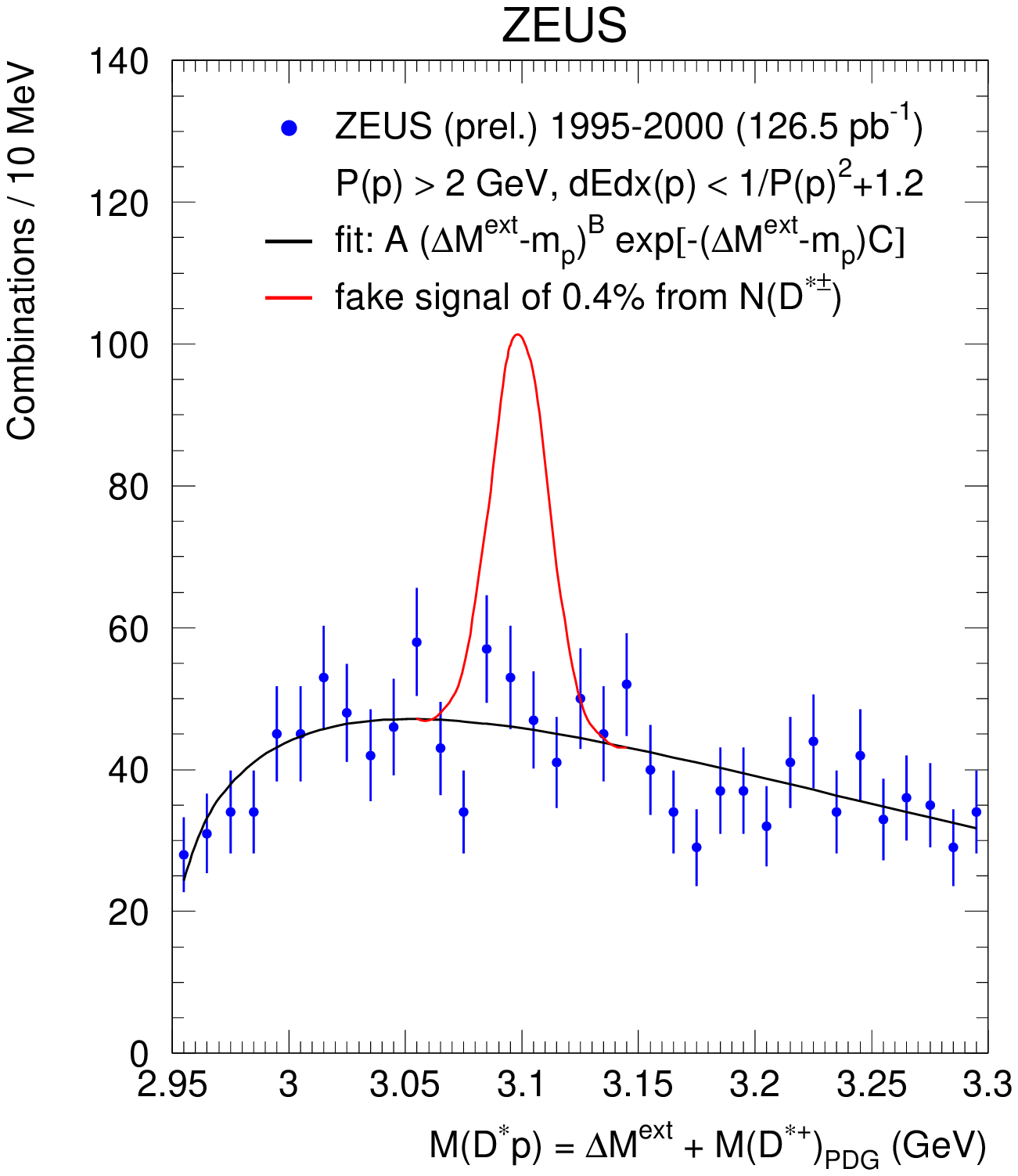}}
\caption[*]{
The distribution of $M(D^{*\pm}p^\mp)$
obtained by the ZEUS collaboration
with low-momentum (left) and high-momentum (right) proton selections.
}
\end{center}
\label{fig:zeusch5q}
\end{figure}

Recent results on the excited charmed-strange meson studies
have been reported by the Belle collaboration~\cite{belle}.
The BABAR collaboration discovery of the narrow $D_s(2317)$ meson
decaying to $D_s\pi^0$~\cite{babar_ds} was confirmed
by the CLEO collaboration~\cite{cleo_ds}. The CLEO collaboration was also
established the state $D_s(2457)$
decaying to $D^*_s\pi^0$~\cite{cleo_ds}.
The Belle collaboration confirmed both new excited charmed-strange mesons
and measured the following $D_s(2457)$ relative branching ratios:
$$B(D_s(2457)\rightarrow D_s\gamma)/
B(D_s(2457)\rightarrow D^*_s\pi^0)=0.55\pm0.13\pm0.08,$$
$$B(D_s(2457)\rightarrow D_s\pi^+\pi^-)/
B(D_s(2457)\rightarrow D^*_s\pi^0)=0.14\pm0.04\pm0.02.$$
Upper limits on the branching ratios
$B(D_s(2317)\rightarrow D_s\gamma$),
$B(D_s(2317)\rightarrow D^*_s\gamma$),
$B(D_s(2317)\rightarrow D_s\pi^+\pi^-$),
$B(D_s(2457)\rightarrow D^*_s\gamma$) and
$B(D_s(2457)\rightarrow D_s\pi^0$) were also set.
These results and an analysis of the helicity distributions
of the $D_s(2317)$ and $D_s(2457)$ mesons in fully reconstructed
$B$ decays permitted to constrain the spin-parity of
these mesons to $0^+$ for $D_s(2317)$ and
$1^+$ for $D_s(2457)$.

The Belle collaboration has reported~\cite{belle} measurements of the excited
neutral charmed mesons with the following masses and widths:
$$M(D_0^{*0})=2308\pm17\pm15\pm28\,{\rm MeV},
~\Gamma(D_0^{*0})=276\pm21\pm18\pm60\,{\rm MeV};$$
$$M(D_1^{\prime 0})=2427\pm26\pm20\pm15\,{\rm MeV},
~\Gamma(D_1^{\prime 0})=384^{+107}_{-75}\pm24\pm70\,{\rm MeV}.$$

The Belle Collaboration discovery of the narrow state $X(3872)$
decaying to $J/\psi \pi^+\pi^-$~\cite{belle_3872} has been confirmed by the
BABAR~\cite{babar},
D0~\cite{d0} and
CDF~\cite{cdf} collaborations.

\section{$D/B$ decays}

The flavour-changing neutral current decays
$D^0\rightarrow\mu^+\mu^-$ and $B_{d,s}^0\rightarrow\mu^+\mu^-$
are strongly suppressed in the Standard Model (SM).
Anomalously large branching ratios of the decays would
indicate new physics beyond the SM, e.g. R-parity violating SUSY.

In the $D$-meson sector, the most stringent published upper limit,
$\mathcal{B}(D^0\rightarrow\mu^+\mu^-)<2.5\times10^{-6}$ (90\% C.L.),
was set by the CDF Collaboration~\cite{cdfd0mumu,donati}.
The HERA-B Collaboration has reported~\cite{herabd0mumu}
results of a search for the $D^0\rightarrow\mu^+\mu^-$ decay
with the data collected during the 2002-2003 HERA running period.
The measurement was based on normalising the number of events
in the $D^0$ signal region to the number of reconstructed
$J/\psi\rightarrow\mu^+\mu^-$ events. Using the $D^0$ and
$J/\psi$ production cross sections measured elsewhere,
the upper limit was obtained to be
$\mathcal{B}(D^0\rightarrow\mu^+\mu^-)<2.0\times10^{-6}$ (90\% C.L.).

The CDF Collaboration has reported results of a search
for $B_{d,s}^0\rightarrow\mu^+\mu^-$ decays with $171\,$pb$^{-1}$
of Run II data~\cite{donati}.
The upper limits were obtained to be
$\mathcal{B}(B^0_d\rightarrow\mu^+\mu^-)<1.5\times10^{-7}$ (90\% C.L.)
and
$\mathcal{B}(B^0_s\rightarrow\mu^+\mu^-)<5.8\times10^{-7}$ (90\% C.L.).
The D0 Collaboration has reported
results of a search for the $B_{s}^0\rightarrow\mu^+\mu^-$ decay
with $240\,$pb$^{-1}$ of Run II data~\cite{lehner}.
Using reconstructed $B^\pm\rightarrow J/\psi K^\pm$ events for normalisation,
the upper limit was set to be
$\mathcal{B}(B^0_s\rightarrow\mu^+\mu^-)<4.6\times10^{-7}$ (95\% C.L.).
The obtained upper limits constrain several SUSY models.

\section{Top production}

Progress on the $t$-quark mass and production cross section
measurements
has been reported by the CDF Collaboration~\cite{lysak}.
Top production cross section is expected to be $30-40\%$ larger at
Run II due to the 
increase of the Tevatron centre-of-mass energy from $1.8\,$TeV to
$1.96\,$TeV.
The top pair production cross section was measured using dilepton,
lepton+jets and all-hadronic modes. The measured cross sections
are in agreement with each other and with the Run I results.
The NLO QCD predictions for the top production cross section
describe the experimental results well. No evidence for single top production
was found.

\section{Beauty production}

The high performance of the upgraded CDF detector has been demonstrated
by the new beauty production cross section measurements~\cite{cdf}.
Using a new dimuon trigger, CDF recorded central $J/\psi\rightarrow\mu^+\mu^-$
production down to $p_T=0$.
The fraction of $J/\psi$ arising from $B$-meson decays was unfolded
using the displaced vertex identification with the help of
the new silicon vertex tracker.
\begin{figure}[!ht]
\vspace*{0.0cm}
\begin{center}
\centerline{\epsfxsize=2.8in\epsfbox{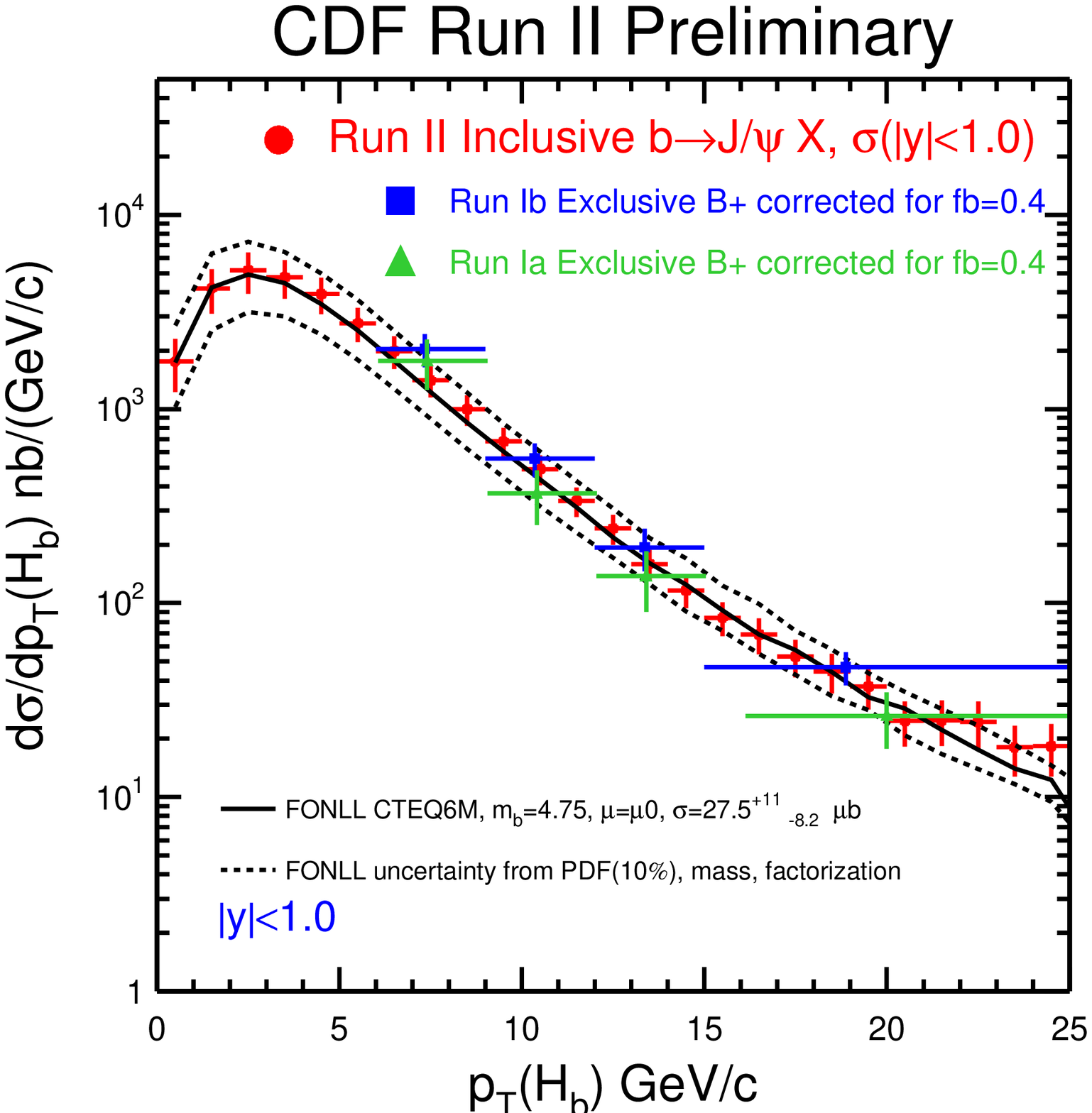}
\hspace*{-0.5cm}
\epsfxsize=2.7in\epsfbox{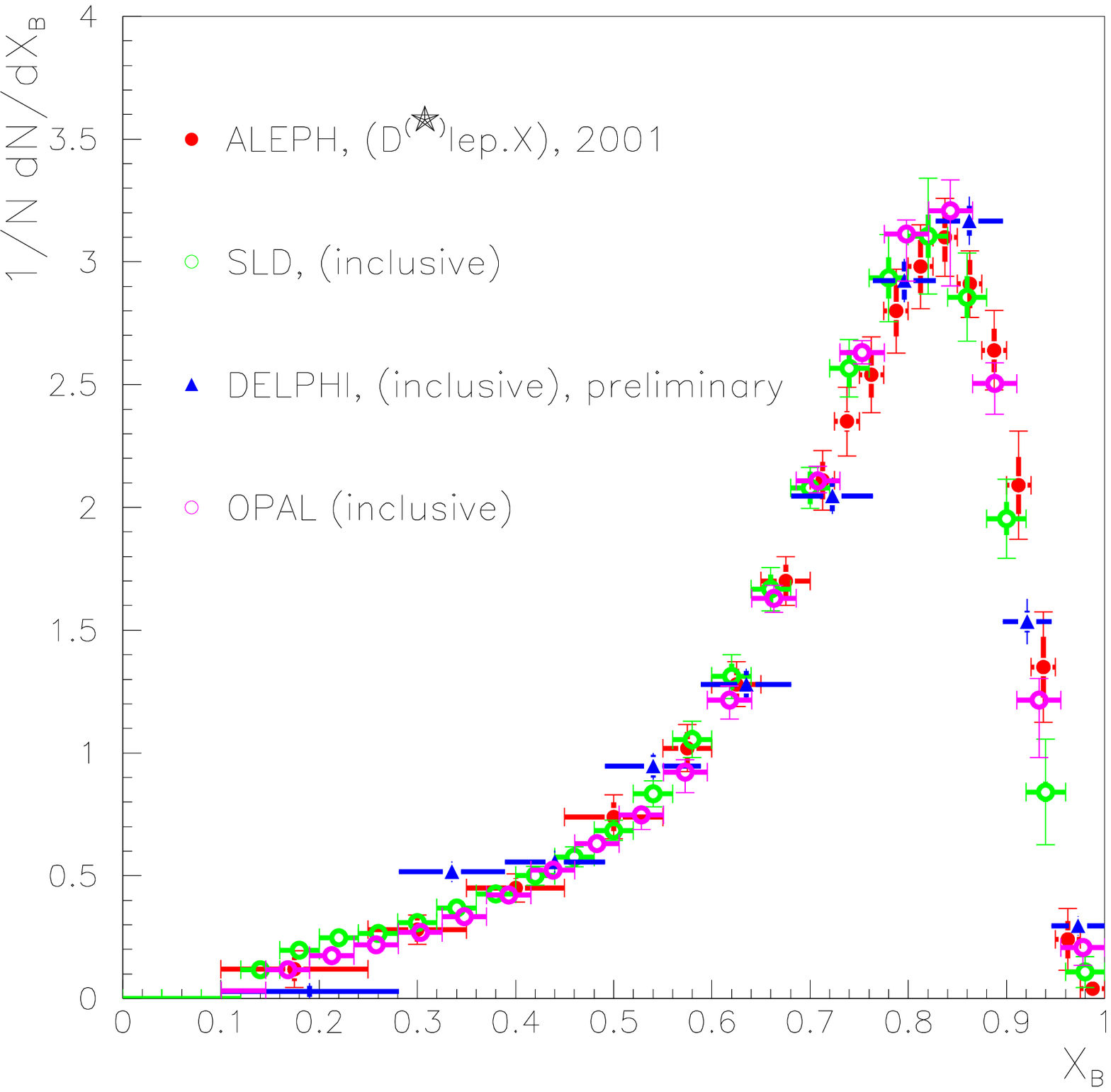}}
\caption[*]{
(Left) Differential cross section for inclusive B hadron production
as a function of $p_T$ measured by CDF.
(Right) the measured distribution for $x_B$ from ALEPH, DELPHI,
OPAL nd SLD. 
}
\end{center}
\label{fig:cdfb}
\end{figure}
The resulting differential cross section for inclusive B hadron production
as a function of $p_T$ is shown in Fig.~3(left).
The measured cross sections agree with the Run I measurements.
The data are well described by predictions
of the matched FONLL calculations~\cite{fonll}
and the MC@NLO Monte Carlo~\cite{frixione}.
The improved agreement between the measured and predicted beauty
hadroproduction cross sections is connected in part with the description
of the b quark hadronisation.

To constrain the b quark hadronisation parameters, the LEP and SLD
measurements were used~\cite{weiser}.
Fig.~3(right) shows the LEP and SLD measurements of the $1/N dN/x_B$
distribution, where $x_B=2E_B/\sqrt{s}$ and $E_B$ is the $b$-hadron energy. 
The measurements favour the LUND and Bowler fragmentation models
and disfavour the Peterson model.
To obtain the b-hadron production fractions,
the LEP and TEVATRON measurements were combined producing:
$f_{B_u}=f_{B_d}=(39.8\pm1.0)\%$,
$f_{B_s}=(10.5\pm1.5)\%$
$f_{b-baryon}=(9.9\pm1.7)\%$.

A new $b$-quark mass measurement has been reported by the DELPHI
Collaboration~\cite{costa}.
Using the 3-jet rate, the running mass was measured to be
$$m_b(M_Z)=2.85^{+0.18}_{-0.19}({\rm stat})\pm 0.13({\rm exp})
^{+0.19}_{-0.20}({\rm had})\pm 0.12({\rm theor})\,{\rm GeV}.$$
A compatible result was obtained using 4-jet events.

The 4-jet rate in $e^+e^-$ collisions is sensitive to the $b$ quark
production by gluon splitting ($g\rightarrow b\bar{b}$).
The fraction of hadronic $Z^0$ decays with $g\rightarrow b\bar{b}$
was obtained~\cite{giammanco} by combining LEP and SLD measurements
to be $(2.54\pm0.51)\times10^{-3}$.
The result is well described by the theoretical predictions
at next-to-leading logarithmic approximation~\cite{gqqtheor}.

A new measurement of the beauty production in proton-nucleus
collisions at $\sqrt{s}\approx41.6\,$GeV has been reported by
the HERA-B Collaboration~\cite{masciocchi}.
Using $35\%$ of the $J/\psi$ sample collected in 2002-2003,
the preliminary result is
$$\sigma(b{\bar b})=12.3^{+3.5}_{-3.2}({\rm stat})\,{\rm nb/nucleon}$$
with the systematic uncertainty of around $25\%$.
The new result is $1.5\,\sigma$ lower
the previous HERA-B measurement~\cite{herab_bb}
taking into account uncertainties of the both measurements.

New results on beauty production with a muon
in the final state in $ep$ scattering have been reported
by the H1~\cite{meyer} and
ZEUS~\cite{wichmann,turcato} Collaborations.
The beauty signals were extracted using the transverse
momentum distribution of the muon relative to the axis of the
associated jet and, in the H1 analyses, the impact parameter distribution
of muons.
The measured cross sections were compared
to the fixed-order (``massive'') NLO QCD predictions
and various Monte Carlo models.
\begin{figure}[!ht]
\vspace*{0.0cm}
\begin{center}
\centerline{\epsfxsize=3.8in\epsfbox{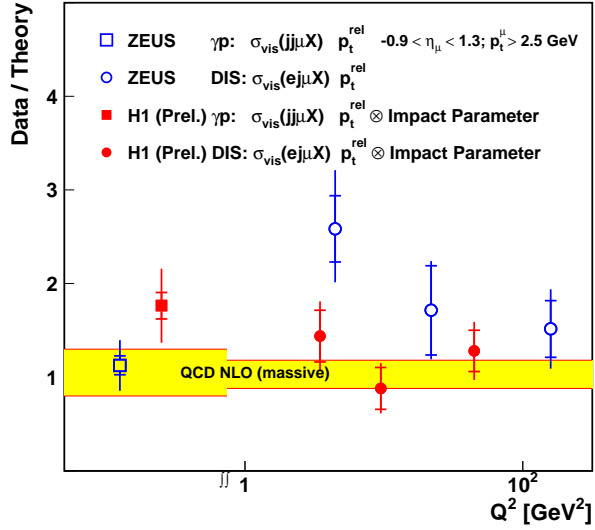}}
\caption[*]{
Ratio of measured $b$-production cross sections at HERA to NLO QCD
expectations as function of $Q^2$.
}
\end{center}
\label{fig:herabb}
\end{figure}
Fig.~4 shows the ratio of the recent HERA measurements
of $b$-production cross sections to the fixed-order NLO QCD expectations
as function of $Q^2$.
The NLO predictions are generally smaller than but agree with the data within
large experimental and theoretical uncertainties.
Shapes of various differential beauty production cross sections
are generally well reproduced by the NLO predictions.
The HERA analyses reported somewhat stronger
disagreements with the NLO predictions at low values of muon
transverse momentum and low values of $Q^2$ and Bjorken $x$ in DIS.

\section{Open charm production}

The CDF Collaboration has reported the inclusive
charm production cross section measurement~\cite{cdf,cdfc}.
The differential cross sections of
$D^0$, $D^{*\pm}$, $D^\pm$ and $D_s^\pm$ production were measured as a function
of transverse momentum in the central rapidity range $|y|<1$.
Both FONLL calculations~\cite{fonll} and ``massless'' calculations from
Kniehl et al. describe shapes
of the measured distributions. The absolute normalisation of the measured
cross sections is slightly underestimated by the predictions.

New measurement of inclusive jet cross sections in $D^{*\pm}$ photoproduction
has been reported by the ZEUS Collaboration~\cite{kohno}.
Differential cross sections as a function of $E_T^{\rm jet}$ and
$\eta^{\rm jet}$ were compared to the fixed-order NLO QCD predictions.
The correction from parton-level to hadron-level jets was done by multiplying
the NLO QCD prediction by hadronisation correction factors,
evaluated using PYTHIA and HERWIG Monte Carlo simulations.
The predictions underestimate the data normalisation but agree with that
within large theoretical uncertainties.
The shapes of the $d\sigma /d E_T^{\rm jet}$ distributions are well
described by the NLO predictions. There is an indication
of a stronger data underestimation by the predictions at the very
highest $E_T^{\rm jet}$.
The shapes of the $d\sigma /d \eta^{\rm jet}$ distributions are described
by the NLO predictions after applying the hadronisation correction.

The high-statistics measurement of charm production in DIS has been
reported by the ZEUS Collaboration~\cite{wing}.
The differential $D^{*\pm}$ cross sections as a function of
$Q^2$, $x$, $p_T(D^{*\pm})$ and $\eta(D^{*\pm})$ are reasonably well
described by the fixed-order NLO predictions.
The predictions using the ZEUS NLO fit gives a better description than that
using CTEQ5F3 or GRV98-HO for the cross section $d\sigma /d \eta(D^{*\pm})$.
The double-differential cross section in $y$ and $Q^2$ was used to extract
the open-charm contribution, $F_2^{c{\bar c}}$,
to the proton structure function $F_2$.

Both charm and beauty contribution to the proton structure function
have been measured by the H1 Collaboration
for $Q^2>110\,$GeV$^2$~\cite{thompson}.
In this analysis,
all events with at least 1 reconstructed track with hits from the central
silicon vertex tracker were used.
The transverse distance of closest approach of the track to the primary
vertex was used to separate the different quark flavours.
Fig.~5 shows the relative charm (H1, ZEUS) and beauty (H1)
contributions to the total $ep$ cross section
as a function of $x$ for two $Q^2$ values.
The data are well described by the predictions from
the H1 NLO QCD fit in which the $c$ and $b$ quarks are teated in
the ``massless'' scheme.

\begin{figure}[!ht]
\vspace*{-0.2cm}
\begin{center}
\centerline{\epsfxsize=4.0in\epsfbox{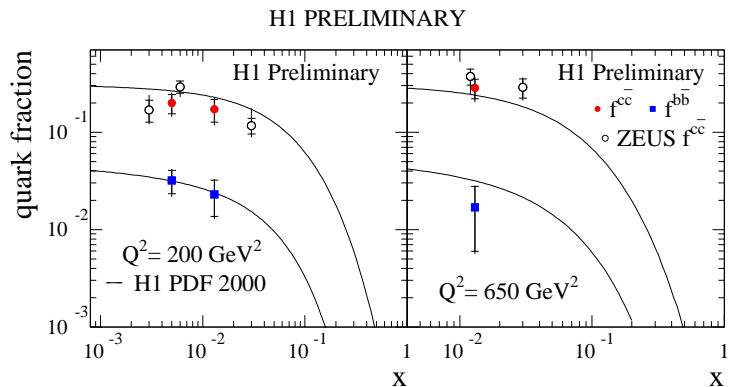}}
\caption[*]{
The relative charm (H1, ZEUS) and beauty (H1) contributions
to the total $ep$ cross section as a function of $x$ for two $Q^2$ values.
The predictions of the H1 NLO QCD fit are also shown.
}
\end{center}
\label{fig:f2bb}
\end{figure}

The ZEUS Collaboration has measured the open-charm contribution,
$F_2^{D(3),c{\bar c}}$,
to the diffractive proton structure function~\cite{vlasov}.
The diffractive charm production was identified by reconstructing
of $D^{*\pm}$ mesons in DIS events with a large rapidity gap
between a proton at high rapidities and the centrally-produced
hadronic system.
The measurement was performed in the diffractive kinematic range
$x_{\scriptscriptstyle {I\!\!P}}<0.35$ and $\beta<0.8$.
Fig.~6 shows
the quantity $x_{\scriptscriptstyle {I\!\!P}}F_2^{D(3),c{\bar c}}$
as a function of $\log(\beta)$ for different $Q^2$ and
$x_{\scriptscriptstyle {I\!\!P}}$ values.
The curves show the NLO ACTW calculations with proton diffractive structure
function fits B, D and SG~\cite{actw}.
Only the fit B predictions agree with the data.

\begin{figure}[!ht]
\vspace*{0.0cm}
\begin{center}
\centerline{\epsfxsize=3.5in\epsfbox{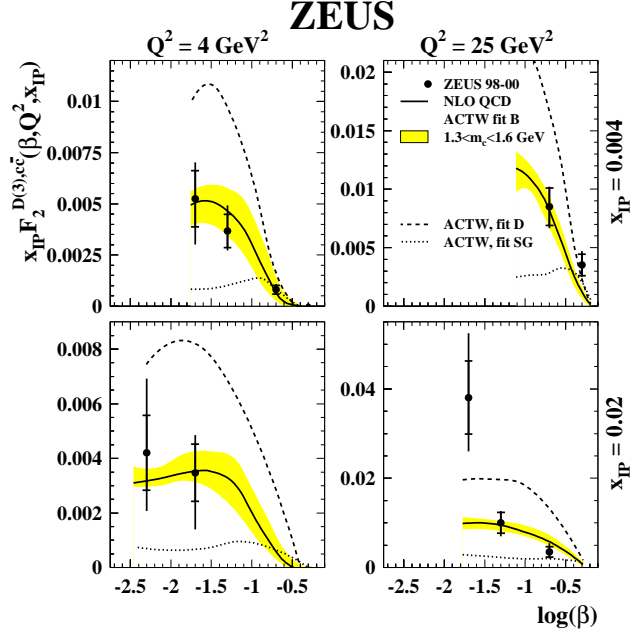}}
\caption[*]{
The quantity $x_{\scriptscriptstyle {I\!\!P}}F_2^{D(3),c{\bar c}}$
as a function of $\log(\beta)$ for different $Q^2$ and
$x_{\scriptscriptstyle {I\!\!P}}$ values.
The curves show the NLO ACTW calculations with proton diffractive structure
function fits B, D and SG.
}
\end{center}
\label{fig:f2dif}
\end{figure}

\section{Charmonium production}

The $J/\psi$ hadroproduction cross section, measured by
the CDF Collaboration~\cite{cdf} down to $p_T(J/\psi)=0$, is more than
one order larger than the charmonium cross sections measured at
Run I for $p_T(J/\psi)$ values above $4-5\,GeV$.
New resummed colour octet matrix elements are awaited
to perform a comparison of the new measurement with the NRQCD predictions.

The HERA-B Collaboration has reported new measurement of charmonium production
cross sections in proton-nucleus collisions, differential distributions
and nuclear effects~\cite{staric}. Using the power law for $J/\psi$ cross
section in $pA$ collisions,
$\sigma_{pA}=\sigma_{pN}A^{\alpha(x_F)}$, the suppression parameter
$\alpha$ was measured down to $x_F=-0.3$.
Using the full accumulated statistics, the measurement will be able
to validate various theoretical predictions.

The ZEUS Collaboration has performed new measurement of the $J/\psi$
helicity distributions using the full data sample from HERA I~\cite{bertolin}.
The results are described by the NRQCD predictions using both colour singlet
and colour octet contributions. Using only the colour singlet contribution,
the predictions are somewhat above the data.

The Belle Collaboration has reported an update of their measurement
of double charmonium production in $e^+e^-$ annihilations\cite{uglov}.
The spectrum of recoil masses against the $J/\psi$ was fitted using
known charmonium states. The measured cross sections are well above of
the HQET predictions. One can conclude that the double charmonium production
in $e^+e^-$ annihilations is not yet understood.
The production cross sections and helicity
characteristics of processes $e^+e^-\rightarrow D^{(*)}{\bar D}^{(*)}$,
observed by the Belle Collaboration for the first time, are generally
described by the theoretical predictions.

\section{Neutrino charm production}

A wide spectrum of results on charm physics in neutrino-induced processes
has been reported by the CHORUS collaboration~\cite{topaksu}.
The inclusive and quasi-elastic $\Lambda_c$ production was measured in
$\nu_\mu$ charged-current interactions at an average neutrino energy
of $27\,$GeV. The quasi-elastic component of the $\Lambda_c$ production
cross section was found to be $\approx 15\%$.
The charm contribution to the cross section of charged-current interactions
induced by ${\bar \nu}_\mu$ was also measured.

New results on the strange sea asymmetry have been reported by the NuTeV
Collaboration~\cite{nutev}.
The dimuon data were fitted in LO and NLO QCD using different methods
of parameterising the strange sea. The fits tended toward a small
negative asymmetry at low $x$, always consistent with zero.
Using the same data, the CTEQ group has extracted a positive
asymmetry. NuTeV and CTEQ will continue collaborate to resolve possible
discrepancies in the theoretical assumptions used.
Both NuTeV and CTEQ agree that
the NuTeV $\sin^2\theta_W$ discrepancy~\cite{nutevsin}
is unlikely covered completely by
an asymmetry in the strange and
anti-strange PDFs.

\section{Conclusions}

The charm hadron spectroscopy is an actual topic again.
The discovery $D_s(2317)$ by the BABAR Collaboration
and $X(3872)$ by the Belle Collaboration were  confirmed
by other experiments.
The situation with pentaquarks in general and, in particular, with
the charmed pentaquark is not clear. More experimental efforts are needed
to clarify the picture.

New results on top, beauty and charm production provides additional
opportunities for testing
the perturbative QCD predictions. In general, the pQCD
calculations can describe the experimental results.
In many cases, the predictions are below the data that is often hidden
by large experimental and theoretical uncertainties.
Increasing experimental precision of the Tevatron and HERA data
on charm and beauty production requires more precise theoretical
calculations. One should add that
there is still no adequate theoretical description of
the double charmonium production in $e^+e^-$ annihilations.

The neutrino experiments provide invaluable information
on charm production and on the strange component of the proton structure.
Recent NuTeV analysis
reports no significant asymmetry
between strange and anti-strange PDFs.

\section*{Acknowledgements}

I thank all participants for their talks and participation in discussions.
I also thank Karin Daum and Stefan Kretzer for co-convening this
working group.  
Special thanks are due to the DIS04 organisers for their help
in the session organisation.

\end{document}